\documentclass[a4paper]{article}
\usepackage{spconf,amsmath,graphicx,subfigure}
\usepackage{booktabs}
\usepackage{multirow}
\usepackage{threeparttable}
\usepackage{url}

\makeatletter
\def\hlinew#1{%
  \noalign{\ifnum0=`}\fi\hrule \@height #1 \futurelet
   \reserved@a\@xhline}
\makeatother

\title{DST: Deformable Speech Transformer for Emotion Recognition}

\name{Weidong Chen$^1$ \qquad Xiaofen Xing$^1$\sthanks{Corresponding author: Xiaofen Xing, xfxing@scut.edu.cn} \qquad Xiangmin Xu$^{23}$ \qquad Jianxin Pang$^4$  \qquad Lan Du$^5$}

\address{$^1$ School of Electronic and Information Engineering, South China University of Technology, China \\
$^2$ School of Future Technology, South China University
of Technology, China \\$^3$ Pazhou Laboratory, China \qquad $^4$ UBTECH Robotics Corp, China  \qquad  $^5$ iFLYTEK Research, China}

\begin{document}

\maketitle
\begin{abstract}
Enabled by multi-head self-attention, Transformer has exhibited remarkable results in speech emotion recognition (SER). Compared to the original full attention mechanism, window-based attention is more effective in learning fine-grained features while greatly reducing model redundancy. However, emotional cues are present in a multi-granularity manner such that the pre-defined fixed window can severely degrade the model flexibility. In addition, it is difficult to obtain the optimal window settings manually. In this paper, we propose a \textbf{D}eformable \textbf{S}peech \textbf{T}ransformer, named DST, for SER task. DST determines the usage of window sizes conditioned on input speech via a light-weight decision network. Meanwhile, data-dependent offsets derived from acoustic features are utilized to adjust the positions of the attention windows, allowing DST to adaptively discover and attend to the valuable information embedded in the speech. Extensive experiments on IEMOCAP and MELD demonstrate the superiority of DST. 

\end{abstract}

\begin{keywords}
speech emotion recognition, deformable network, Transformer, deformable attention mechanism
\end{keywords}

\section{Introduction}
\label{sec:intro}
Emotion is one of the most essential characteristics that distinguishes humans from robots \cite{intro} and speech is the most basic tool for daily communication \cite{perception}.
Therefore, analyzing emotion states through speech signals is a continuing concern for the research community. Owing to the rapid development of deep learning, many advanced models have been proposed and delivered promising results in speech emotion recognition (SER). In particular, convolutional neural networks \cite{ISNet,audio_cnn}, recurrent neural networks \cite{SER_RNN1,co_attention} and their variants \cite{DECN,gru_attention,graph_meld} have been widely studied and deployed for applications.

\begin{figure}[t]
  \centering
  \includegraphics[width=0.95\linewidth]{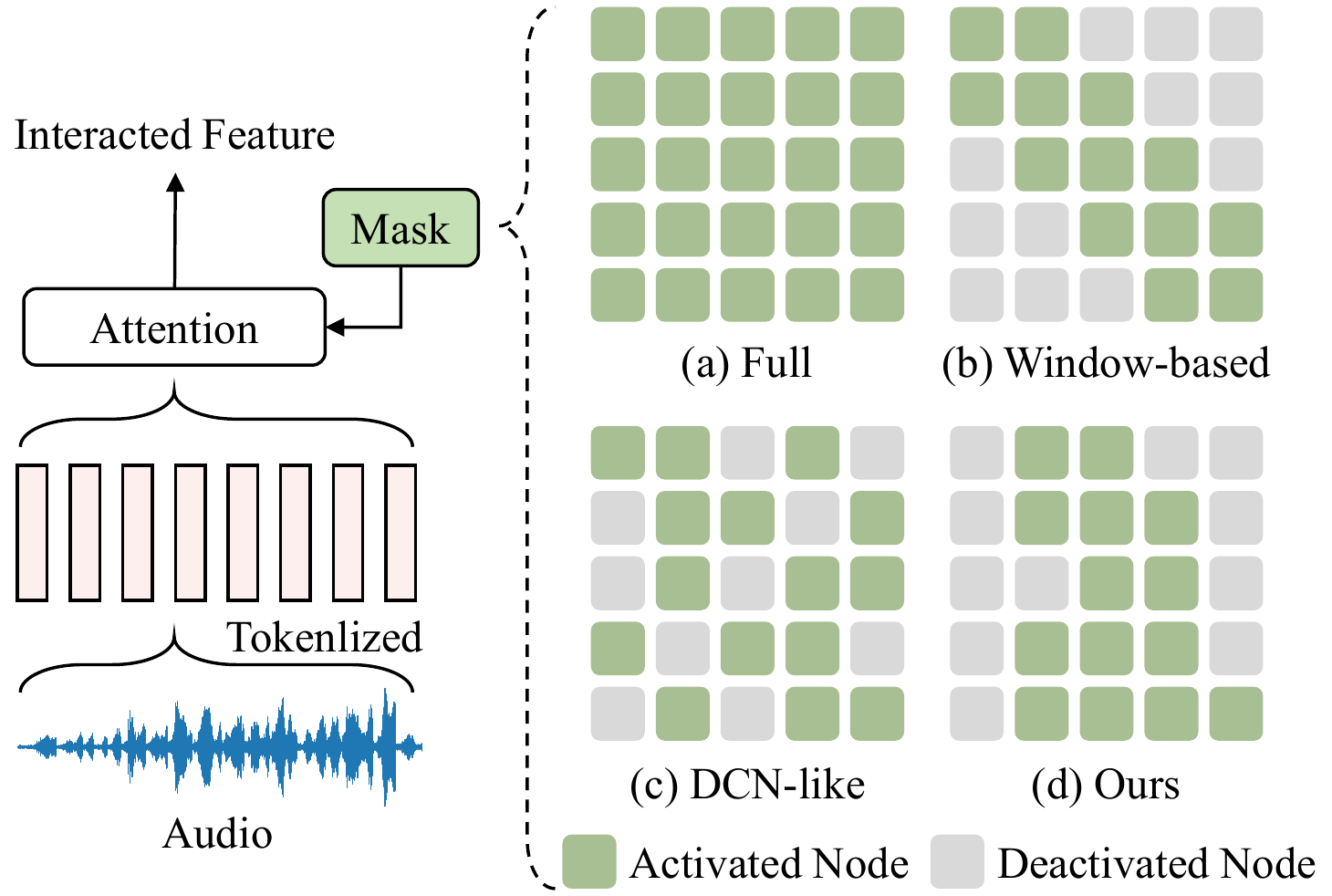}
  \caption{Comparison between different attention mechanisms. In contrast to prior works that have pre-set window sizes or fixed window positions, we propose to make them both flexible and deformable. DCN-like attention is applied in vision.} 
  \label{fig:introduction}
\end{figure}

Transformer \cite{transformer}, which is the recent white hope architecture, is making a splash in deep learning domain. 
Different from previous networks, Transformer adopts the full attention mechanism, which is depicted in Fig.~\ref{fig:introduction}(a), to learn a global representation of input signal.
Although the effectiveness of Transformer in SER has already been confirmed \cite{speechformer,ctnet,ksT}, there are several key points to be aware of when handling emotion analysis with Transformer: 1) Emotional cues are multi-grained in nature, which means that beyond the global representation, the details in speech are also important. For example, the local characteristics, such as articulation and prolongation, are highly relevant to the emotion states.
2) The full attention mechanism suffers from a lack of diversity and is thus inadequate to capture the multi-granularity features.
3) The computation of the full attention is quite redundant.

One mainstream approach to improve Transformer is employing the window-based attention mechanism \cite{speechformer,audio_trans}. As shown in Fig~\ref{fig:introduction}(b), window-based attention restricts the attention scope to a fixed local window whose size is typically set to a small value to focus on the fine-grained features. However, the immutable window also severely decreases the flexibility of model. What is worse, it weakens the ability of global learning and thus it inevitably requires considerable manual tuning of window configuration to obtain peak performance.

\begin{figure*}[t]
  \centering
  \includegraphics[width=0.96\linewidth]{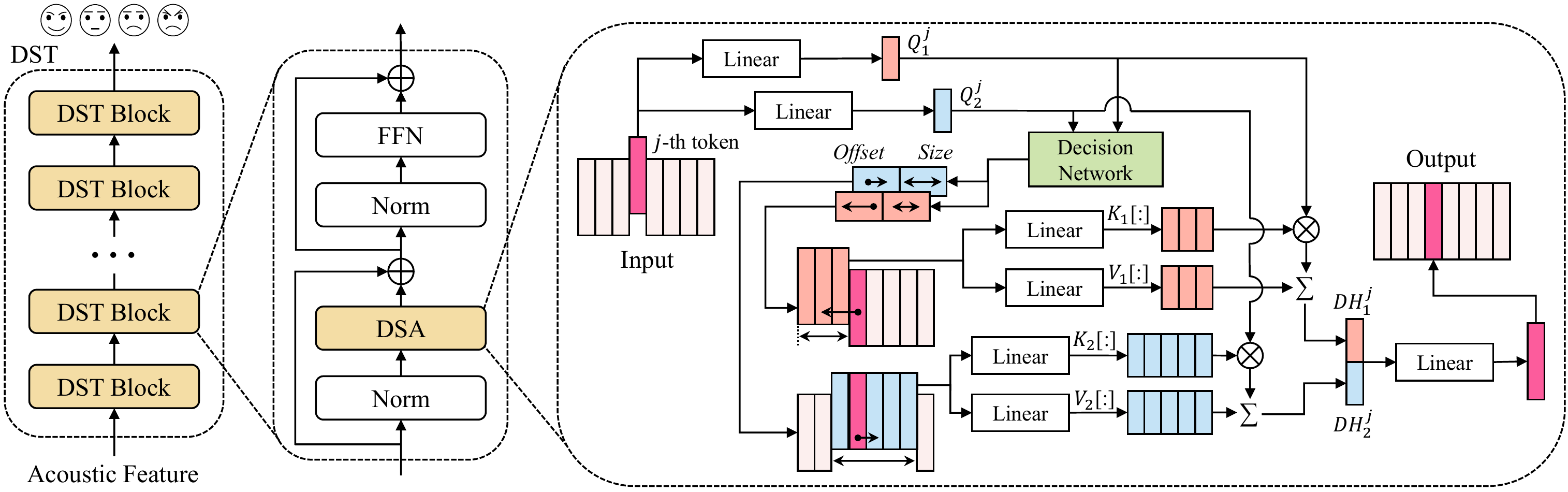}
  \caption{Overview structure of the proposed DST. The only difference between DST block and the vanilla Transformer is the replacement of MSA with DSA. In DSA, we show only $h=2$ attention heads and omit the softmax operation for a clear presentation. $\otimes$ and $\oplus$ represent the matrix multiplication and addition, respectively. $\Sigma$ represents the weighted summation.}
  \label{fig:framework}
\end{figure*}

To alleviate the above issues, this paper proposes a deformable framework, named DST, for speech emotion recognition. In DST, the window sizes are learned by a light-weight decision network based on the input speech, breaking the limitations of using the pre-set configuration. Also, the window positions can be shifted by learned offsets on a per-input basis. These qualities follow the natures of emotion and greatly improve the model flexibility. In addition, unlike deformable convolutional networks (DCNs) \cite{dcn} and DCN-like attention utilized in vision that model intermittently (Fig.~\ref{fig:introduction}(c)) \cite{dcn_vit_1,dcn_vit_2}, DST models continuous tokens (Fig.~\ref{fig:introduction}(d)), which is more in line with the continuous speech signal.
Finally, we visualize different attention mechanisms for an intuitive understanding.
The contributions of this work are summarized as follows: 
\begin{itemize}
\item We endow Transformer with deformability by employing flexible, data-dependent window sizes and offsets.
\item Extensive experiments on IEMOCAP \cite{IEMOCAP} and MELD \cite{meld} datasets show that DST outperforms the state-of-the-art approaches. Our codes are publicly available at \url{https://github.com/HappyColor/DST}.
\end{itemize}

\section{Methodology}
The proposed DST, as illustrated in Fig.~\ref{fig:framework}, is composed of multiple stacked DST blocks. Each DST block mainly consists of a deformable speech attention (DSA) module and a feed-forward network (FFN). Equipped with the DSA module, the system is able to adaptively determine the usage of window sizes and window positions depending on the input speech signal, which greatly improves the model flexibility and can learn the multi-granularity emotional cues effectively.

\subsection{Revisiting Transformer}

At the core of the standard Transformer is the multi-head self-attention module (MSA), which makes Transformer stand out from other deep neural networks. 
More details can be found in \cite{transformer}.
Specifically, the MSA mechanism can be written as:
\begin{equation}
  \emph{\textbf{Q}}_i = \emph{\textbf{Q}}\emph{\textbf{W}}_i^Q\,,\  \emph{\textbf{K}}_i = \emph{\textbf{K}}\emph{\textbf{W}}_i^K\,,\  \emph{\textbf{V}}_i = \emph{\textbf{V}}\emph{\textbf{W}}_i^V
  \label{eq1}
\end{equation}
\begin{equation}
  \emph{\textbf{H}}_i = softmax(\frac{\emph{\textbf{Q}}_i\emph{\textbf{K}}_i^\top}{\sqrt{d_{Q}}})\emph{\textbf{V}}_i
  \label{eq2}
\end{equation}
\begin{equation}
  MSA(\emph{\textbf{Q}},\emph{\textbf{K}},\emph{\textbf{V}}) = concat(\emph{\textbf{H}}_1,..., \emph{\textbf{H}}_h)\emph{\textbf{W}}^o
  \label{eq3}
\end{equation}
where \emph{\textbf{Q}}, \emph{\textbf{K}}, \emph{\textbf{V}} are query, key and value matrices, respectively; $d_{Q}$ is a scaling factor and $h$ denotes the number of attention heads; $\emph{\textbf{W}}_i^Q$, $\emph{\textbf{W}}_i^K$, $\emph{\textbf{W}}_i^V$ and $\emph{\textbf{W}}^o$ are to be learned parameters.

\subsection{Deformable Speech Transformer}

\subsubsection{Deformable Speech Attention}
Deformable speech attention (DSA) is at the core of the DST. 
Different from previous attention mechanisms, DSA is able to change the window sizes and modify the window positions via a simple decision network.
Let $\emph{\textbf{Q}}_i^j$ be the $j$-th token of $\emph{\textbf{Q}}_i$ in the $i$-th attention head, where $i\in[1,h]$. The decision network first produces the window size $s_{ij}$ and offset $o_{ij}$ conditioned on $\emph{\textbf{Q}}_i^j$:
\begin{equation}
  \bar{s}_{ij}, \bar{o}_{ij} = \emph{\textbf{Q}}_{i}^j\emph{\textbf{W}}_i^D
  \label{eq4}
\end{equation}
\begin{equation}
  s_{ij} = \sigma_1(\bar{s}_{ij})\times L\,,\  o_{ij} = \sigma_2(\bar{o}_{ij})\times L
  \label{eq5}
\end{equation}
where $j\in [0,L-1]$ and $L$ denotes the sequence length of the features; $\emph{\textbf{W}}_i^D$ is the parameter matrix; $\sigma_1$ and $\sigma_2$ are two nonlinear functions for restricting the range of the outputs. For example, the window size $s_{ij}$ should lie in the range $(0,L)$. Therefore, we first apply the sigmoid function to limit the value of $\bar{s}$ to $(0,1)$ and then scale it by the maximum length $L$. Similarly, since the valuable information can be on either side of the current $j$-th token, we apply the tanh function to normalize $\bar{o}_{ij}$ to the range $(-1,1)$ before scaling it by $L$.

Giving the current position index $j$ and the offset $o_{ij}$, the anchor of the critical segment $A_{ij}$ can be obtained. Combining with the predicted window size $s_{ij}$, the boundaries of the attention window for the $j$-th query in the $i$-th head, $L_{ij}$ and $R_{ij}$, are also given. The calculations are as follows:
\begin{equation}
  A_{ij} = j + o_{ij}
  \label{eq6}
\end{equation}
\begin{equation}
  L_{ij} = A_{ij} - s_{ij}\,,\  R_{ij} = A_{ij} + s_{ij}
  \label{eq7}
\end{equation}

Finally, each query token attends to its respective deformed attention windows through the proposed DSA mechanism.
The DSA is formulated as follows:
\begin{equation}
  \emph{\textbf{DH}}_{i}^j = softmax(\frac{\emph{\textbf{Q}}_{i}^j\emph{\textbf{K}}_i[L_{ij}:R_{ij}]^\top}{\sqrt{d_{Q}}})\emph{\textbf{V}}_i[L_{ij}:R_{ij}]
  \label{eq8}
\end{equation}
\begin{equation}
  DSA(\emph{\textbf{Q}},\emph{\textbf{K}},\emph{\textbf{V}}) = concat(\emph{\textbf{DH}}_{1},..., \emph{\textbf{DH}}_{h})\emph{\textbf{W}}^o
  \label{eq9}
\end{equation}
where $\emph{\textbf{K}}_i[L_{ij}:R_{ij}]$ and $\emph{\textbf{V}}_i[L_{ij}:R_{ij}]$ consist of the $L_{ij}$-th to the $R_{ij}$-th tokens of $\emph{\textbf{K}}_i$ and $\emph{\textbf{V}}_i$ matrices, respectively; $\emph{\textbf{DH}}_{i}^j$ denotes the $j$-th output token of the $i$-th attention head.

\subsubsection{End-to-End Training}
\begin{figure}[t]
  \centering
  \includegraphics[width=\linewidth]{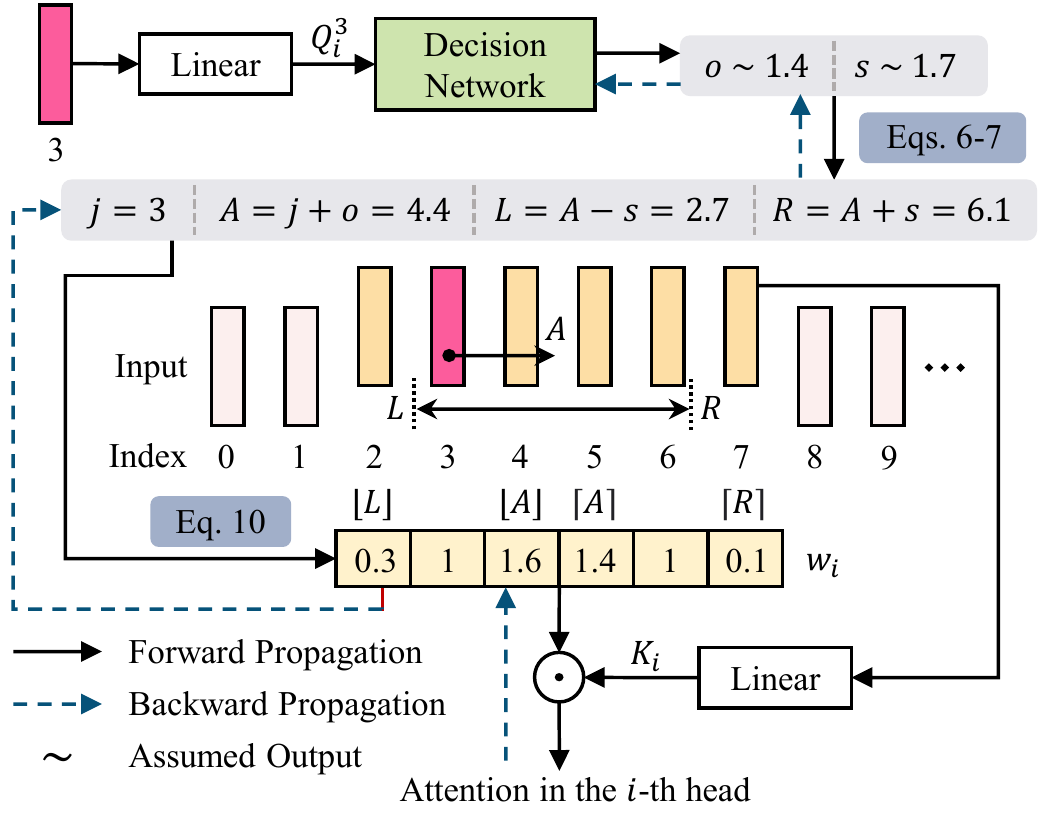}
  \caption{Differentiable weighting process for end-to-end training. $\odot$ denotes position-wise multiplication.}
  \label{fig:backwards}
\end{figure}
For ease of reading, we will omit the subscripts of the notations.
In practice, the outputs of the decision network, the window size $s$ and offset $o$, are decimals, causing the attention boundaries, $L$ and $R$, to be decimals as well.
However, in Eq.~\ref{eq8}, the indexing operations $K[L:R]$ and $V[L:R]$ require both $L$ and $R$ to be integers. One simple solution is rounding the $L$ and $R$ to integers $\lfloor L \rfloor$ and $\lceil R \rceil$, where $\lceil \cdot \rceil$ and $\lfloor \cdot \rfloor$ round a number up and down, respectively. However, the rounding operations are non-differentiable, resulting in a decision network that cannot be optimized by the back propagation algorithm.
To add the decision network to the computation graph in a differentiable way,
we leverage the distances between the predicted boundaries ($L$ and $R$) and the true boundaries ($\lfloor L \rfloor$ and $\lceil R \rceil$), and the distances between the central tokens ($\lfloor A \rfloor$ and $\lceil A \rceil$) and the anchor ($A$) to yield weights for the selected key features in DSA. 
In general, only if the predicted boundaries are close to the true boundaries, the $\lfloor L \rfloor$-th and the $\lceil R \rceil$-th tokens will be assigned large weights. The weights for two central tokens are against each other, and whichever side the anchor is close to has a larger weight. Noting that we expect anchor to be the center of the important segment, thus the weights for the central tokens should be larger than 1 to emphasize them. Overall, the weights are computed as below:
\begin{equation}
  w^k_i=
  \begin{cases}
    1-(L-\lfloor L \rfloor)  & \text{if}\  k=\lfloor L \rfloor \\ 
    1-(\lceil R \rceil-R)  & \text{if}\  k=\lceil R \rceil \\ 
    1+(\lceil A \rceil-A)  & \text{if}\  k=\lfloor A \rfloor \\ 
    1+(A-\lfloor A \rfloor)  & \text{if}\  k=\lceil A \rceil \\
    1  & \text{otherwise}
  \end{cases}
  \label{eq10}
\end{equation}
where $k\in [\lfloor L \rfloor, \lceil R \rceil]$ denotes the token index and $w^k_i$ is the weight for the $k$-th token in the $\emph{\textbf{K}}_i$ matrix. Eventually, $s$ and $o$ are correlated with the weights, and the process of weighting is differentiable. The decision network can be optimized with the entire model jointly in an end-to-end manner.
Suppose the current index $j$ is 3, the weighting process is shown in Fig.~\ref{fig:backwards}.


\section{Experiments}
\label{sec:exp}

\subsection{Datasets and Acoustic Features}
\textbf{IEMOCAP} \cite{IEMOCAP} contains five sessions, every of which has one male and one female speaker, respectively. 
We merge excitement into happiness category and select 5,531 utterances from happy, angry, sad and neutral classes. Experiments are conducted in leave-one-session-out cross-validation strategy. 
\textbf{MELD} \cite{meld} dataset contains 13,708 utterances with 7 emotion classes. 
As MELD has been officially split into training, validation and testing sets, we use the validation set for hyper-parameter turning and report the scores on the testing set. 
To be consistent with previous works, weighted accuracy (WA), unweighted accuracy (UA) and weighted average F1 (WF1) are used to assess the model performance.

\noindent \textbf{Features.} Pre-trained self-supervised WavLM \cite{wavlm} is adopted to extract the acoustic features. 
The max sequence lengths are set to 326 and 224 for IEMOCAP and MELD, respectively.

\subsection{Training Details and Hyper-Parameters}
The number of training epochs is set to 120. SGD \cite{sgd} with a learning rate of $5e^{-4}$ on IEMOCAP and $1e^{-3}$ on MELD is applied to optimize the model. Cosine annealing warm restarts scheduler \cite{scheduler} is used to adjust the learning rate in the training phase. Learning rate of the decision network is multiplied by a factor of 0.1. The batch size is 32. The number of attention heads is 8. The number of DST blocks is 4.

\subsection{Experimental Results and Analysis}

\subsubsection{Comparison with Other Attention Mechanisms}
\label{sec:exp_attn}
\begin{figure}[t]
  \centering
  \includegraphics[width=0.95\linewidth]{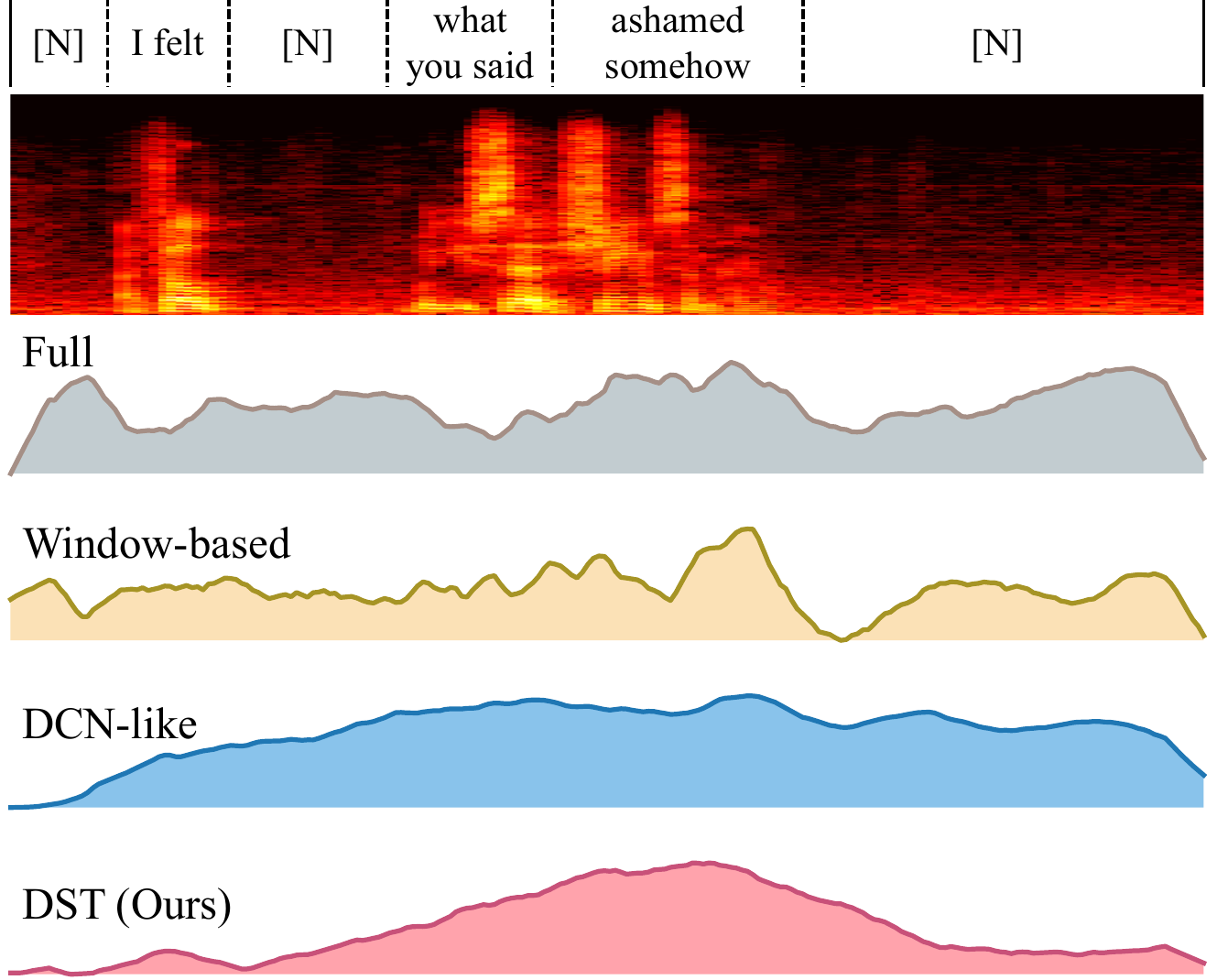}
  \caption{Visualization of different attention mechanisms. [N] indicates a silent segment with background noise.}
  \label{fig:visualization}
\end{figure}

\textbf{Performance Analysis.} To analyze the potency of DST, we implement other common attention mechanisms, namely, full \cite{transformer}, window-based and DCN-like \cite{dcn_vit_1} attentions, for comparison. The fixed window size of the window-based attention and the number of sampling points in the DCN-like attention are empirically set to 10\% of input length. 
Average percentage of activated tokens for each query is also listed for comprehensive analysis. As shown in Table~\ref{tab_1}, DST outperforms the counterparts on IEMOCAP and MELD by a considerable margin. In particular, the use of DCN-like attention causes a significant drop in performance, which means that modeling the continuous tokens is essential for speech signal. Most interestingly, we find that on IEMOCAP, each query attends to an average of 8.7\% of all input tokens, while on MELD, this percentage increases to 12.7\%. This ambiguity exposes the difficulty of manual tuning and recommends configurations that are automatically determined by model itself.
Also, we find that DST can learn all potential emotional features, both fine and coarse, through its deformable capabilities.
Furthermore, we discard the learned window size ($-deform.$ size) or reset the offset to zero ($-deform.$ offset), and the ablation results shown in the last two rows of Table~\ref{tab_1} once again confirm the effectiveness of the proposed deformable design.

\noindent \textbf{Visualization Analysis.}
To further understand the proposed model, we consider an utterance sample and intuitively compare the attention weights in each attention mechanism by visualization. As illustrated in Fig~\ref{fig:visualization}, voiced fragments are distributed in a small part of the entire speech sample. The full attention has difficulty in highlighting the key parts owing to the large amount of noise that deeply confuses the model. Although the window-based attention is able to learn the fine-grained features, it is inevitably limited in performance when the duration and position of the key segments mismatch with the pre-defined window settings. Since speech is a continuous signal, the DCN-like attention fails to judge importance by the discrete tokens, leading to little difference in its assigned weights. Inspiringly, DST successfully focuses on the critical segments (\textit{``ashamed somehow'' in text}) and highlights them by means of the learned window sizes and offsets.

\subsubsection{Comparison to Previous State-of-the-Art}

Table~\ref{tab_2} gives the comparison among the proposed DST with some known approaches on IEMOCAP and MELD. All approaches here adopt acoustic features as input for a fair comparison. On IEMOCAP, DST outperforms the previous best results obtained by \cite{ISNet,co_attention}. On MELD, DST substantially surpasses the other competitors by a considerable margin.

\begin{table}[t]
    \caption{Performances of adopting different attention mechanisms on two corpora. PAT indicates the average percentage of the activated tokens for each query during the testing phase. $-deform.$ \textup{x} means the corresponding attribute \textup{x} is fixed.}
    \label{tab_1}
    \centering
    \begin{threeparttable}
    \begin{tabular}{cccccc}
    \hline
    \multirow{2}{*}[-3pt]{\shortstack{Attention\\Mechanisms}}   & \multicolumn{3}{c}{\multirow{1}{*}[-2pt]{IEMOCAP}}  & \multicolumn{2}{c}{\multirow{1}{*}[-2pt]{MELD}}    \\ \cmidrule(lr){2-4} \cmidrule(lr){5-6}
                 & WA    & UA  & PAT  & WF1  & PAT \\ \hline
    Full         & 0.710 & 0.720 & 100 & 0.472 & 100 \\
    Window-based & 0.714 & 0.723 & 10  & 0.476 & 10  \\
    DCN-like     & 0.665 & 0.681 & 10  & 0.455 & 10  \\  \hline
    DST (Ours)         & \textbf{0.718} & \textbf{0.736} & 8.7 & \textbf{0.488} & 12.7 \\ 
    $-deform.$ size  & 0.714 & 0.726 & 10  & 0.479 & 10 \\
    $-deform.$ offset & 0.716 & 0.729 & 8.9  & 0.483 & 12.8\\  \hline
    \end{tabular}
    \end{threeparttable}
\end{table}

\begin{table}[t]
    \caption{Comparison with known state-of-the-art systems on IEMOCAP and MELD. All systems apply audio as input.}
    \label{tab_2}
    \centering
    \begin{threeparttable}
    \begin{tabular}{cc||cc}
    \hline
    \multicolumn{4}{c}{\multirow{1}{*}[-1pt]{IEMOCAP}}  \\ \hline
    Method & Year & WA & UA \\ \hline
    Audio-CNN \cite{audio_cnn} & 2021 & 0.654 & 0.667 \\
    AR-GRU \cite{gru_attention}         & 2021 & 0.669 & 0.683 \\
    ISNet \cite{ISNet}      & 2022 & 0.704 & 0.650 \\
    Co-attention \cite{co_attention}  & 2022 & 0.698 & 0.711 \\  \hline
    DST (Ours)  & 2023 & \textbf{0.718} & \textbf{0.736} \\ \hline \hline
    \multicolumn{4}{c}{\multirow{1}{*}[-1pt]{MELD}}  \\ \hline
    Method & Year & \multicolumn{2}{c}{WF1} \\ \hline
    CTNet \cite{ctnet}         & 2021 & \multicolumn{2}{c}{0.382} \\
    DECN \cite{DECN}        & 2021 & \multicolumn{2}{c}{0.439} \\
    SpeechFormer \cite{speechformer} & 2022 & \multicolumn{2}{c}{0.419} \\
    MM-DFN \cite{graph_meld}      & 2022 & \multicolumn{2}{c}{0.427} \\ \hline
    DST (Ours)  & 2023 & \multicolumn{2}{c}{\textbf{0.488}} \\ \hline
    \end{tabular}
    \end{threeparttable}
\end{table}

\section{Conclusion}
In this paper, a deformable speech Transformer, named DST, has been proposed for speech emotion recognition. DST can capture the multi-granularity emotional cues effectively via deformed attention windows whose sizes and positions are automatically determined by model itself. This deformability of DST significantly improves the model flexibility and adaptability.
Experimental results on IEMOCAP and MELD corpora demonstrate the effectiveness of the proposed DST. We hope our work can inspire insights towards designing flexible and potent variants of Transformer for the speech domain. 
In the future, we plan to extend DST to other speech tasks and further verify its adaptability.

\section{Acknowledgement}
The work is supported in part by the Natural Science Foundation of Guangdong Province 2022A1515011588, in part by the National Key R\&D Program of China under Grant 2022YFB4500600, in part by the Science and Technology Project of Guangzhou under Grant 202103010002, in part by the Science and Technology Project of Guangdong under Grant 2022B0101010003, in part by the National Natural Science Foundation of China under Grant U1801262, and in part by the Guangdong Provincial Key Laboratory of Human Digital Twin  under Grant 2022B1212010004.

\bibliographystyle{IEEEbib}
\bibliography{refs}

\begin{thebibliography}{10}

\bibitem{intro}
J.-M. Fellous,
\newblock ``From human emotions to robot emotions,''
\newblock {\em Architectures for Modeling Emotion: Cross-Disciplinary
  Foundations, American Association for Artificial Intelligence}, pp. 39--46,
  2004.

\bibitem{perception}
B.C. Moore, L.K. Tyler, and W.~Marslen-Wilson,
\newblock ``Introduction. the perception of speech: from sound to meaning,''
\newblock {\em Philosophical transactions of the Royal Society of London.
  Series B, Biological sciences}, vol. 363, no. 1493, pp. 917--921, 2008.

\bibitem{ISNet}
W.~Fan, X.~Xu, B.~Cai, and X.~Xing,
\newblock ``Isnet: Individual standardization network for speech emotion
  recognition,''
\newblock {\em IEEE/ACM Transactions on Audio, Speech, and Language
  Processing}, vol. 30, pp. 1803--1814, 2022.

\bibitem{audio_cnn}
Z.~Peng, Y.~Lu, S.~Pan, and Y.~Liu,
\newblock ``Efficient speech emotion recognition using multi-scale cnn and
  attention,''
\newblock in {\em IEEE International Conference on Acoustics, Speech and Signal
  Processing}, 2021, pp. 3020--3024.

\bibitem{SER_RNN1}
J.~Lee and I.~Tashev,
\newblock ``High-level feature representation using recurrent neural network
  for speech emotion recognition,''
\newblock in {\em Proc. Interspeech}, 2015, pp. 1537--1540.

\bibitem{co_attention}
H.~Zou, Y.~Si, C.~Chen, D.~Rajan, and E.~S. Chng,
\newblock ``Speech emotion recognition with co-attention based multi-level
  acoustic information,''
\newblock in {\em IEEE International Conference on Acoustics, Speech and Signal
  Processing}, 2022, pp. 7367--7371.

\bibitem{DECN}
Z.~Lian, B.~Liu, and J.~Tao,
\newblock ``Decn: Dialogical emotion correction network for conversational
  emotion recognition,''
\newblock {\em Neurocomputing}, vol. 454, pp. 483--495, 2021.

\bibitem{gru_attention}
S.~T. Rajamani, K.~T. Rajamani, A.~Mallol-Ragolta, S.~Liu, and B.~Schuller,
\newblock ``A novel attention-based gated recurrent unit and its efficacy in
  speech emotion recognition,''
\newblock in {\em IEEE International Conference on Acoustics, Speech and Signal
  Processing}, 2021, pp. 6294--6298.

\bibitem{graph_meld}
D.~Hu, X.~Hou, L.~Wei, L.~Jiang, and Y.~Mo,
\newblock ``Mm-dfn: Multimodal dynamic fusion network for emotion recognition
  in conversations,''
\newblock in {\em IEEE International Conference on Acoustics, Speech and Signal
  Processing}, 2022, pp. 7037--7041.

\bibitem{transformer}
A.~Vaswani, N.~Shazeer, N.~Parmar, J.~Uszkoreit, L.~Jones, and et~al.,
\newblock ``Attention is all you need,''
\newblock in {\em Proceedings of the 31st International Conference on Neural
  Information Processing Systems}, 2017, pp. 5998--6008.

\bibitem{speechformer}
W.~Chen, X.~Xing, X.~Xu, J.~Pang, and L.~Du,
\newblock ``{SpeechFormer: A Hierarchical Efficient Framework Incorporating the
  Characteristics of Speech},''
\newblock in {\em Proc. Interspeech}, 2022, pp. 346--350.

\bibitem{ctnet}
Z.~Lian, B.~Liu, and J.~Tao,
\newblock ``Ctnet: Conversational transformer network for emotion
  recognition,''
\newblock {\em IEEE/ACM Transactions on Audio, Speech, and Language
  Processing}, vol. 29, pp. 985--1000, 2021.

\bibitem{ksT}
W.~Chen, X.~Xing, X.~Xu, J.~Yang, and J.~Pang,
\newblock ``Key-sparse transformer for multimodal speech emotion recognition,''
\newblock in {\em IEEE International Conference on Acoustics, Speech and Signal
  Processing}, 2022, pp. 6897--6901.

\bibitem{audio_trans}
K.~Chen, X.~Du, B.~Zhu, Z.~Ma, T.~Berg-Kirkpatrick, and S.~Dubnov,
\newblock ``Hts-at: A hierarchical token-semantic audio transformer for sound
  classification and detection,''
\newblock in {\em IEEE International Conference on Acoustics, Speech and Signal
  Processing}, 2022, pp. 646--650.

\bibitem{dcn}
J.~Dai, H.~Qi, Y.~Xiong, Y.~Li, G.~Zhang, H.~Hu, and Y.~Wei,
\newblock ``Deformable convolutional networks,''
\newblock in {\em Proceedings of the IEEE International Conference on Computer
  Vision}, 2017, pp. 764--773.

\bibitem{dcn_vit_1}
Z.~Xia, X.~Pan, S.~Song, L.~E. Li, and G.~Huang,
\newblock ``Vision transformer with deformable attention,''
\newblock in {\em Proceedings of the IEEE/CVF Conference on Computer Vision and
  Pattern Recognition}, 2022, pp. 4794--4803.

\bibitem{dcn_vit_2}
X.~Zhu, W.~Su, L.~Lu, B.~Li, X.~Wang, and J.~Dai,
\newblock ``Deformable detr: Deformable transformers for end-to-end object
  detection,''
\newblock {\em arXiv preprint arXiv:2010.04159}, 2020.

\bibitem{IEMOCAP}
C.~Busso, M.~Bulut, C.-C Lee, A.~Kazemzadeh, E.~Mower, and et~al.,
\newblock ``Iemocap: Interactive emotional dyadic motion capture database,''
\newblock {\em Language Resources and Evaluation}, vol. 42, no. 4, pp.
  335--359, 2008.

\bibitem{meld}
S.~Poria, D.~Hazarika, N.~Majumder, G.~Naik, E.~Cambria, and R.~Mihalcea,
\newblock ``{MELD: A} multimodal multi-party dataset for emotion recognition in
  conversations,''
\newblock {\em arXiv preprint arXiv:1810.02508}, 2019.

\bibitem{wavlm}
S.~Chen, C.~Wang, Z.~Chen, Y.~Wu, S.~Liu, Z.~Chen, J.~Li, N.~Kanda,
  T.~Yoshioka, X.~Xiao, J.~Wu, L.~Zhou, and et~al.,
\newblock ``Wavlm: Large-scale self-supervised pre-training for full stack
  speech processing,''
\newblock {\em IEEE Journal of Selected Topics in Signal Processing}, pp.
  1--14, 2022.

\bibitem{sgd}
H.~Robbins and S.~Monro,
\newblock ``A stochastic approximation method,''
\newblock {\em The annals of mathematical statistics}, pp. 400--407, 1951.

\bibitem{scheduler}
I.~Loshchilov and F.~Hutter,
\newblock ``Sgdr: Stochastic gradient descent with warm restarts,''
\newblock {\em arXiv preprint arXiv:1608.03983}, 2016.

\end{thebibliography}

\end{document}